# Shedding Light on Rechargeable Na/Cl₂ Battery

Guanzhou Zhu[1§], Peng Liang[1§], Cheng-Liang Huang[2,3], Shu-Chi Wu[1], Cheng-Chia Huang[2], Yuan-Yao Li[2], Shi-Kai Jiang[4], Wei-Hsiang Huang[5], Jiachen Li[1], Feifei Wang[6], Bing-Joe Hwang[4], Hongjie Dai[1*]

[1]Department of Chemistry and Bio-X, Stanford University, Stanford, California 94305, USA.

[2]Department of Chemical Engineering, National Chung Cheng University, Chia-Yi 62102, Taiwan.

[3]Department of Electrical Engineering, National Chung Cheng University, Chia-Yi 62102, Taiwan.

[4]Department of Chemical Engineering, National Taiwan University of Science and Technology, Taipei 10607, Taiwan.

[5]National Synchrotron Radiation Research Center (NSRRC), Hsinchu 30076, Taiwan.

[6]Department of Electrical and Electronic Engineering, The University of Hong Kong, Hong Kong 999077, Hong Kong.

§ G.Z. and P.L. contributed equally to this work.

* Email: hdai1@stanford.edu.

**Abstract**

Advancing new ideas of rechargeable batteries represents an important path to meeting the ever increasing energy storage needs. Recently we showed rechargeable sodium/chlorine (Na/Cl₂) (or lithium/chlorine Li/Cl₂) batteries that used a Na (or Li) metal negative electrode, a microporous amorphous carbon nanosphere (aCNS) positive electrode and an electrolyte containing dissolved AlCl₃ and fluoride additives in thionyl chloride (SOCl₂)[1-2]. The main battery redox reaction involved conversion between NaCl and Cl₂ trapped in the carbon positive electrode, delivering a cyclable capacity of up to 1200 mAh g$^{-1}$ (based on positive electrode mass) at a ~ 3.5 V discharge voltage[1-2]. Here, we discovered by X-ray photoelectron spectroscopy (XPS) that upon charging a Na/Cl$_2$ battery, chlorination of carbon in the positive electrode occurred to form C-Cl accompanied by molecular Cl$_2$ infiltrating the porous aCNS, consistent with Cl$_2$ probed by mass spectrometry. Synchrotron X-ray diffraction observed the development of graphitic ordering in the initially amorphous aCNS under battery charging when the carbon matrix was oxidized/chlorinated and infiltrated with Cl$_2$. The C-Cl, Cl$_2$ species and graphitic ordering were reversible upon discharge, accompanied by NaCl formation. The results revealed redox conversion between NaCl and Cl$_2$, reversible graphitic ordering/amorphourization of carbon through battery charge/discharge, and for the first time probed trapped Cl$_2$ in porous carbon by XPS.

**Introduction**

The increasing use of mobile phones, electric vehicles and other devices demands better rechargeable batteries with high capacity/energy density, high safety and long cycle life. Different types of rechargeable battery have been actively pursued using reactive metals as negative electrodes[3-11]. Recently we reported rechargeable $Na/Cl_2$ and $Li/Cl_2$ batteries in an aluminum chloride ($AlCl_3$), fluoride-additive and $SOCl_2$ containing electrolyte[1-2]. Reduction of $SOCl_2$ in the first discharge afforded NaCl or LiCl deposited into the pores and surface of aCNS or defective graphitic carbon in the positive electrode, accompanied by Na or Li oxidation to $Na^+$ or $Li^+$ at the negative electrode. Upon re-charging, NaCl or LiCl was oxidized to $Cl_2$ trapped in the carbon matrix and reduced back to salt upon discharge, thus affording rechargeability for the batteries. Over cycling, redox reactions were complex involving $SOCl_2$ oxidation and regeneration, but the main charge-discharge voltage plateaus were found to be reversible redox between $Na/Na^+$ or $Li/Li^+$ at the negative electrode and between $Cl^-/Cl_2$ at the positive electrode, with a discharge voltage of ~ 3.5 V and cycling capacity of up to 1200 mAh $g^{-1}$ (based on the mass of carbon in the positive electrode) over up to 200 cycles[1-2]. Thus far, we found that amorphous carbon with high surface area/large pore volumes or in situ exfoliated defective graphite were important to high $Cl_2$ trapping capability and rechargeability of alkali metal/$Cl_2$ batteries, with the reversible charge-discharge capacity increasing with the pore volume of carbon[1-2]. Further investigations of alkali metal/$Cl_2$ batteries are key to fundamental understanding of the battery chemistry and devising new strategies to improve battery performance towards practical use.

In this work, we probed aCNS carbon positive electrodes through cycling in $Na/Cl_2$ batteries by several spectroscopy techniques. X-ray photoelectron spectroscopy (XPS) and Auger electron spectroscopy (AES) with scanning Auger nanoprobe mapping indicated that during battery charging, carbon-chlorine (C-Cl) bonds were formed in aCNS and were reversibly cleaved when the battery was later discharged. Synchrotron x-ray diffraction (XRD) results revealed that graphitic ordering appeared in aCNS upon charging and reversed when the battery was discharged. Both XPS and mass spectrometry (MS) revealed $Cl_2$ trapped in charged aCNS electrodes, confirming $Cl_2$ as an important species for battery operation. The trapped $Cl_2$ decreased but remained in the carbon host upon exposure to vacuum for several days, allowing $Cl_2$ probing by XPS for the first time at room temperature.

**Results and Discussions**

Our battery was assembled using a Na metal as the negative electrode, activated aCNS (at 1000 °C in $CO_2$[1, 12-13]) loaded either on Ni foam or SS316 stainless steel mesh as positive electrode, and 4 M $AlCl_3$ dissolved in $SOCl_2$ containing 2 weight percent (2 wt%) of sodium bis(fluorosulfonyl)imide (NaFSI) and sodium bis(trifluoromethanesulfonyl)imide (NaTFSI) additives as the electrolyte (Fig. 1a, Methods)[1]. SEM imaging of aCNS revealed that the material was constituted by nanospheres with an average diameter of ~ 60 nm (Fig. 1a). The Raman spectrum of the as-made aCNS displayed two broad Raman bands, one at ~ 1585 cm$^{-1}$ or the so-called G band, originating from the $E_{2g}$ phonon mode of crystalline graphite, and the other at ~ 1335 cm$^{-1}$ or the so-called D band, originating from the disorder-allowed modes of crystalline graphite[14-15]. The amorphous nature of aCNS was also consistent with XRD spectrum showing no obvious diffraction peak in the 2θ ranging from 20º to 35º (Fig. S1).

We first discharged the battery to 2 V at a current density of 50 mA g$^{-1}$, through which $SOCl_2$ was reduced to form $SO_2$, S and NaCl, with the latter deposited on the aCNS electrode until passivation[1, 16-18]. After the first discharge, the battery was set to cycle at a specific capacity of 1200 mAh g$^{-1}$ with 100 mA g$^{-1}$ current (Fig. 1c). Note that throughout this work, charging was controlled by setting the charging time of the battery at a specific capacity of 1200 mAh g$^{-1}$ (charging time = specific capacity/current), and the discharging was controlled by setting a discharge cutoff voltage of 2 V. We observed a main charging plateau at ~ 3.8 V, corresponding to the oxidation of NaCl to Cl$_2$, and towards the end of charging, a slightly higher plateau at ~ 3.9 V was observed, corresponding to the oxidation of electrolyte in forming SCl$_2$, S$_2$Cl$_2$, and SO$_2$Cl$_2$ (Fig. 1d)[1, 16]. During discharge the main plateau at ~ 3.55 V was due to the reduction of Cl$_2$ back to NaCl, and two extra plateaus, one at ~ 3.7 V and one at ~ 3.18 V, were attributed to reduction of SCl$_2$/S$_2$Cl$_2$ and SO$_2$Cl$_2$, respectively (Fig. 1d)[1, 16].

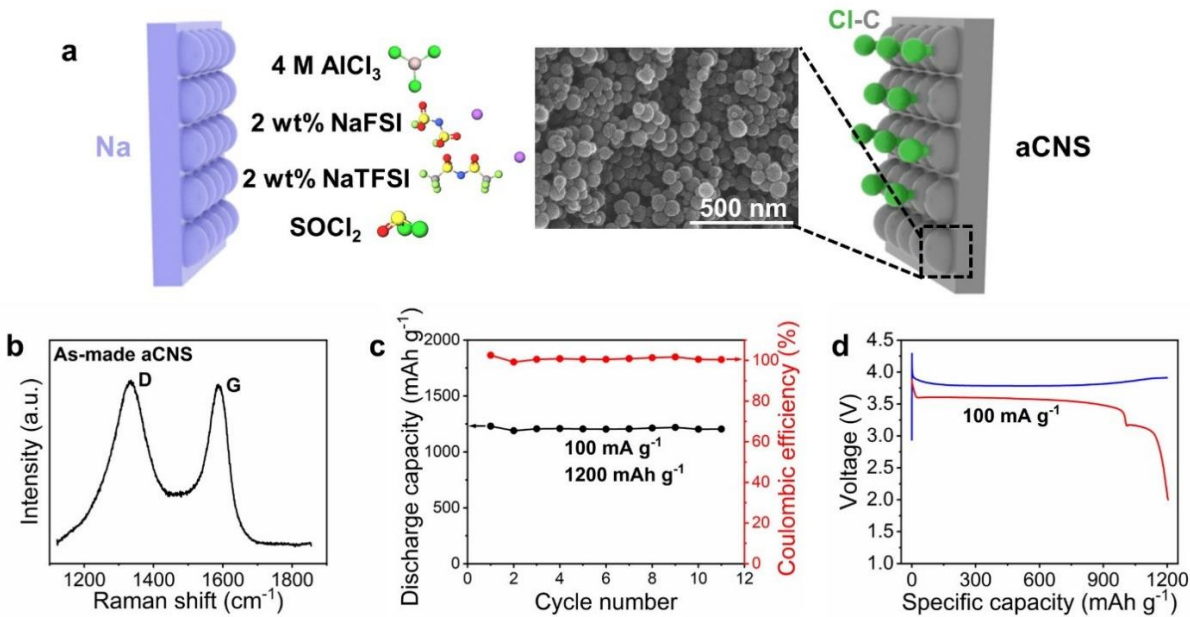

**Figure 1. Schematic drawing of the Na/Cl$_2$ battery, Raman spectrum of the as-made aCNS, and a typical Na/Cl$_2$ battery cycling performance at 1200 mAh g$^{-1}$. A,** Schematic drawing of a Na/Cl$_2$ battery. SEM imaging of aCNS revealed that aCNS was made by nanospheres with an average of ~ 60 nm in diameter. **B,** Raman spectrum of the as-made aCNS, demonstrating amorphous nature with its broad D and G bands. **C,** Cycling performance of a typical Na/Cl$_2$ battery at 1200 mAh g$^{-1}$ with 100 mA g$^{-1}$ current. The battery was stopped at cycle 11 in either charged or discharged state for spectroscopy studies. **D,** Typical charge-discharge

curve of a Na/Cl$_2$ battery at 1200 mAh g$^{-1}$ with 100 mA g$^{-1}$ current. The battery was stopped in either charged or discharged state for spectroscopy studies.

During battery cycling, we stopped the battery and characterized by XPS the charged or discharged aCNS electrode, first by washing with deionized ultra-filtered (DIUF) water (see Methods) to remove all the residual electrolyte and salts (SOCl$_2$, NaCl, AlCl$_3$, NaFSI, etc.). We observed a peak at binding energy of ~ 200 eV (all XPS spectra in this work were calibrated by C 1s=284.8 eV) and attributed it to Cl species in C-Cl bonds (Fig. 2a blue solid curve)[19-20] formed by oxidative chlorination of aCNS during charging. When aCNS was discharged and washed with water, the Cl 2p XPS spectrum displayed a much weaker peak (Fig. 2a red solid curve), indicating that C-Cl bonds formed during battery charging were mostly reversed during discharging. We further annealed the charged/water-washed aCNS in a nitrogen (N$_2$) environment at 600 °C for 45 minutes (Methods) and observed the disappearance of Cl 2p peak at ~ 200 eV (Fig. 2a blue dotted curve), suggesting thermal cleavage of C-Cl bonds[19-21]. The Na 1s and Al 2p XPS spectra of the DIUF-water-washed charged and discharged aCNS electrodes displayed no obvious peak (Fig. S2), indicating that all the Na and Al salts in the electrodes were indeed removed by washing.

The C 1s XPS spectra of DIUF-water-washed charged and discharged aCNS also showed reversible formation and cleavage of C-Cl bonds during battery cycling. When the battery was charged, in addition to the peaks corresponding to sp$^2$ C (~ 284.5 eV), sp$^3$ C (~ 285.0 eV), C-O (~ 286.3 eV), C=O (287.7 eV) and O-C=O (~ 288.7 eV), a new peak, attributed to C-Cl bond formation was present in the C 1s spectrum at ~ 286.6 eV (Fig. 2b top spectrum)[20, 22-24]. Note that oxygen groups in the aCNS were attributed to high-temperature CO$_2$ activation, with the binding energy assignments in agreement with literature data[20, 22-24]. After the battery was discharged, the C-Cl peak at ~ 286.6 eV disappeared in the C 1s spectrum (Fig. 2b bottom spectrum), consistent with the removal of C-Cl bonding upon discharging. Note we also measured the C 1s XPS spectrum of the as-made aCNS and were able to fit it using the same peaks as in the discharged aCNS spectrum (Fig. S3). In charged aCNS the C-Cl peak at ~ 286.6 eV (Fig. 2b) constituted ~ 3.87% of the total C 1s area, and the corresponding chlorination reaction C + Cl$^-$ → C-Cl + $x$ e$^-$ ($x$ < 1 is the charge transfer number) contributed only < ~ 7% to the total 1200 mAh g$^{-1}$ capacity.

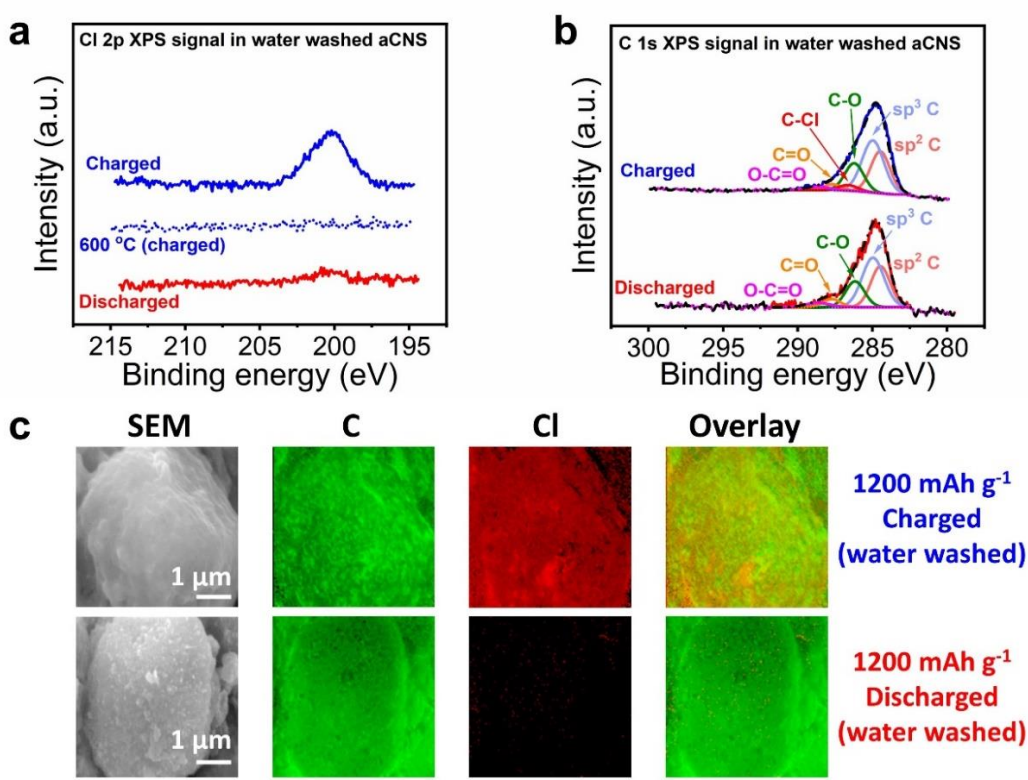

**Figure 2. The formation and cleavage of C-Cl bonds in aCNS during Na/Cl$_2$ battery charge and discharge, respectively. a,** Cl 2p XPS spectra of charged (before and after annealing in N$_2$ at 600 °C) and discharged aCNS electrodes. All aCNS samples were washed using DIUF water. A clear Cl 2p peak at ~ 200 eV was observed in the charged aCNS (before N$_2$ annealing, blue solid curve), indicating the presence of C-Cl bonds. This peak had its intensity decreased to a negligible intensity in discharged aCNS (red solid curve), corresponding to the cleavage of the C-Cl bonds. After annealing in N$_2$ at 600 °C, the peak at ~ 200 eV disappeared (blue dotted curve), suggesting the thermal cleavage of C-Cl bonds during the annealing process. **b,** C 1s XPS spectra of charged and discharged aCNS electrodes after washing with DIUF water. A peak at ~ 286.6 eV, attributed to C-Cl bonds, was present in the charged aCNS spectrum and disappeared in the discharged aCNS spectrum, indicating the formation and cleavage of C-Cl bonds during charging and discharging, respectively. All C 1s XPS fitting was done using the CasaXPS software with a line shape of GL(20) (A line shape constructed by a mixture of Gaussian (80%) and Lorentzian (20%)). All peaks had their binding energy and full width at half maximum (FWHM) restrained to vary by ± 0.1 eV. **c,** AES/SEM mappings of charged and discharged aCNS electrodes after washing with DIUF water. In charged aCNS, C and Cl signals overlapped with each other over aCNS (top row), indicating the presence of C-Cl bonds. In discharged aCNS, only C signal was observed, and no significant Cl signal was detected (bottom row), suggesting the reversible cleavage of the C-Cl bonds.

Auger electron spectroscopy (AES) with a scanning nanoprobe showed that when the aCNS was charged and washed, overlapping C and Cl signal were detected uniformly over aCNS (Fig. 2c, top row), consistent with carbon chlorination with the formation of C-Cl bonds during battery charging. When the aCNS was discharged and washed by water, Cl signal mostly disappeared over aCNS, and only C signal was detected (Fig. 2d, bottom row). These results further confirmed the reversible formation and cleavage of C-Cl bonds during battery charging and discharging, respectively.

Next, we performed synchrotron XRD analysis of the as-made, charged (to 1200 mAh/g), and discharged aCNS electrodes sealed in Ar without any exposure to air or water (see Methods). Two-dimensional XRD ring data showed no obvious diffraction peaks in the as-made aCNS electrode (Fig. 3a), consistent with its highly amorphous nature (Fig. S1). When the aCNS electrode was charged, various new peaks appeared especially in the 2θ range from 20° to 31° (Fig. 3b). The most dominant peaks/rings in this region were at ~ 24.1°, ~ 27.2°, ~ 28.8°, and ~ 31.0° (Fig. 3b). In a discharged aCNS electrode, these peaks were much weaker with negligible intensity (Fig. 3b, c). The 2θ peak at ~ 31.7° for NaCl (200) planes was weaker in charged than discharged electrodes (Fig. 3b, c, d), consistent with electrochemical oxidation of NaCl to $Cl_2$ through charging[1]. A broad XRD peak appeared at ~ 27.2° in the charged aCNS, close to that of graphite (~ 26.7°), suggesting the development of graphitic ordering in the initially amorphous aCNS upon charging. The FWHM of the peak at ~ 27.2° in charged aCNS was ~ 0.139°, corresponding to a graphitic region constituting ~ 180 graphene layers. On discharge, the graphitic peak disappeared and a peak at ~ 27.4° emerged due to the (111) planes of NaCl formed by reduction of trapped $Cl_2$ in aCNS through discharging (Fig. 3d)[25-26]. This was accompanied by an increased NaCl (200) peak in discharged aCNS (Fig. 3b, c, d).

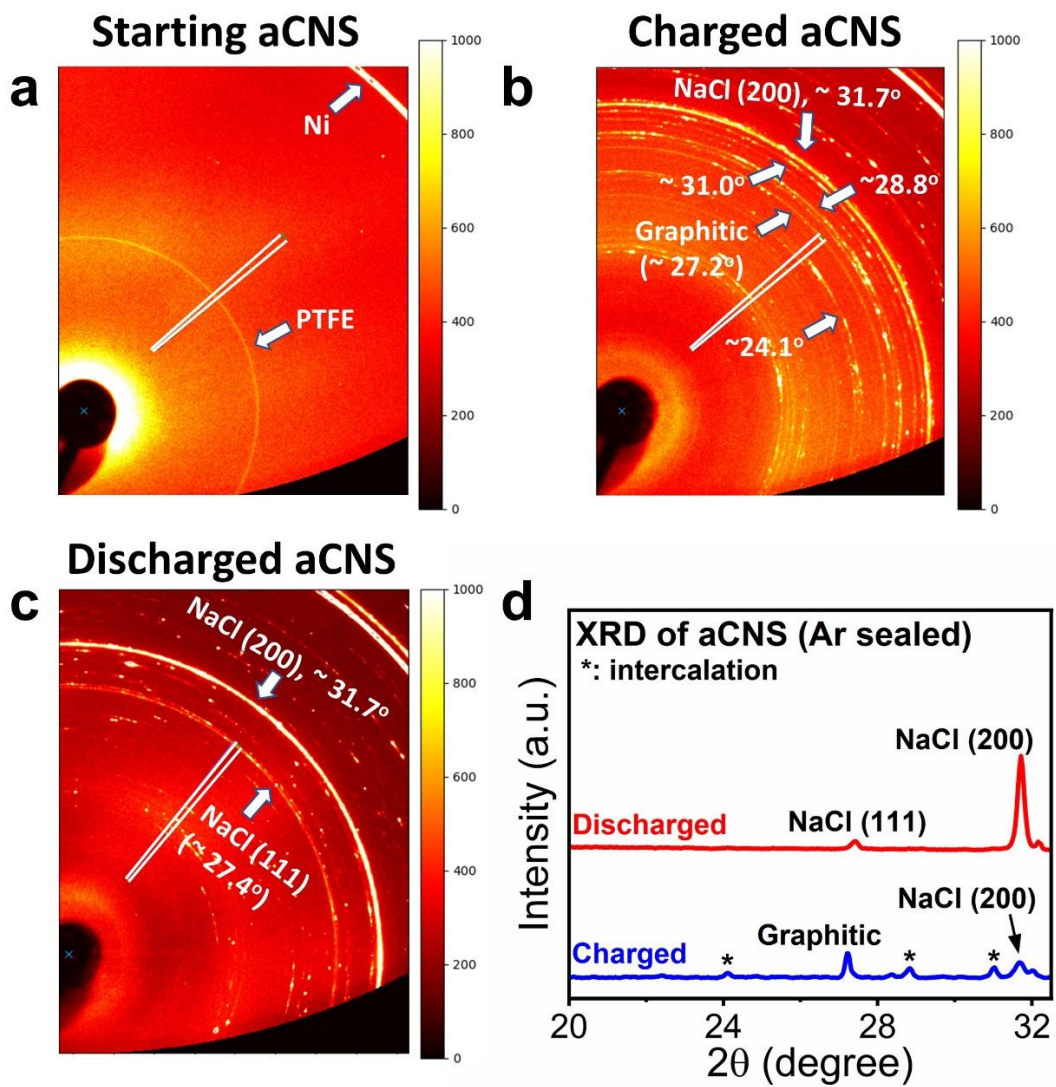

**Figure 3. Two-dimensional XRD ring data and XRD spectra of as-made, charged, and discharged aCNS (without exposing to air or washing by DIUF water). a,** Two-dimensional XRD ring data of as-made aCNS. No obvious peaks were detected in the

range from 20° to 30°, consistent with the amorphous nature of aCNS. The only two peaks present were due to PTFE binder and Ni substrate. The area enclosed by the two white lines indicated the area over which the spectrum presented in Fig. S1 was averaged. **b,** Two-dimensional XRD ring data of charged aCNS. A lot of new peaks were observed in the range from ~ 20° to ~ 31°, corresponding to intercalated graphitic ordering and NaCl. The area enclosed by the two white lines indicated the area over which the spectrum presented in d were averaged. **c,** Two-dimensional XRD ring data of discharged aCNS. Much less peaks were observed in the range from ~ 20° to ~ 31°, with the intensity of the NaCl peak increased. The area enclosed by the two white lines indicated the area over which the spectrum presented in d were averaged. **d,** XRD spectra of charged and discharged aCNS electrodes, averaging over the area enclosed by the two white solid lines in b-c. The peaks labelled with '*' were attributed to intercalated graphitic ordering sites that were developed in charged aCNS. The XRD spectrum presented in this figure had its x-ray wavelength converted to that of copper K-α (1.5406 Å).

Additional peaks appeared in charged aCNS near ~ 24.1°, ~ 28.8°, and ~ 31.0° next to the graphitic peak, attributed to chemical intercalation between graphitic layers in the developed graphitic regions (Fig. 3b, d)[27-28]. These intercalation peaks disappeared when the battery was discharged (Fig. 3c, d), suggesting that the intercalation reactions were reversible. Similar intercalation peaks were observed in our recent studies of Li/Cl$_2$ battery using a defective graphite material as the positive electrode[2]. The intercalating species could involve AlCl$_4^-$·$x$ AlCl$_3$ (where $x$ is a number and AlCl$_3$ was released from NaAlCl$_4$ complexes in the electrolyte upon NaCl electro-oxidization), as suggested in a similar study of electrochemical graphite chlorination and chlorine evolution in a NaAlCl$_4$ melt electrolyte[29-32]. Carbon materials in general upon chlorination/oxidation (as in our case, Fig. 2) became heavily hole-doped and positively charged, allowing facile intercalation of anionic (e.g. chloroaluminate) species[33]. We suggest that upon charging the Cl$_2$ species generated by NaCl electro-oxidation were mostly trapped in the micropores of aCNS, which also facilitated anion intercalation of the graphitic domains adjacent to the pores[1-2]. Related was that chloride intercalation into graphite only occurred when Cl$_2$ was present in various vapor phase reactions[29, 31, 34-35].

Overall, our XPS and XRD results painted a picture that upon charging, the porous carbon nanospheres exhibited chemical and structural changes/evolutions due to chlorination, $Cl_2$ formation from NaCl oxidation and anionic intercalation. The carbon matrix was infiltrated with chlorine in the pores and intercalating species in nanoscale graphitic domains formed due to high mechanical stresses generated electrochemically. Upon discharge, $Cl_2$ was reduced to NaCl mostly filling the nanopores and extended to cover the aCNS surface, accompanied by carbon de-chlorination, graphitic domain de-intercalation and reversion to a more relaxed, amorphous carbon structure. These reactions and processes are reversible over cycling, affording rechargeability of the $Na/Cl_2$ battery over hundreds of cycles. A high performance carbon material should possess sufficient defects/sites for chlorination, large pore volumes/surface areas (in the starting material[1] or after in situ activation[2]) for high $Cl_2$ trapping capacity and mechanical stability.

Next, we performed mass spectrometry measurements of species trapped in aCNS electrodes in fully charged (to 1200 mAh g$^{-1}$) or discharged state using a residual gas analyzer (RGA, Fig. S4), after opening batteries during cycling and sealing the electrodes in an Ar containing chamber without exposure to air. Species evolving from the electrode were continuously vacuum pumped and analyzed over time (Fig. S4, Methods). For charged aCNS electrode, the detected ratio between $Cl_2$ (mass over charge m/z = 70 amu) and $SOCl_2$ (m/z = 118 amu) gradually increased over time, especially after vacuum pumping > 6 h (Fig. 4a, blue curve), whereas the detected $Cl_2/SOCl_2$ ratio for a discharged electrode was much lower and stayed flat over time (Fig. 4a, red curve). This suggested the existence of trapped $Cl_2$ in the charged aCNS electrode (Fig. 4a, b), well exceeding the $Cl_2$ fragments from $SOCl_2$ in the electrolyte residue as in the discharged electrode case (Fig. 4a, c). Similar trend was also observed in $Li/Cl_2$ battery using defective graphite as the positive electrode[2], confirming unambiguously $Cl_2$ trapped in porous carbon as an important reactant for the $Li/Cl_2$ battery operation.

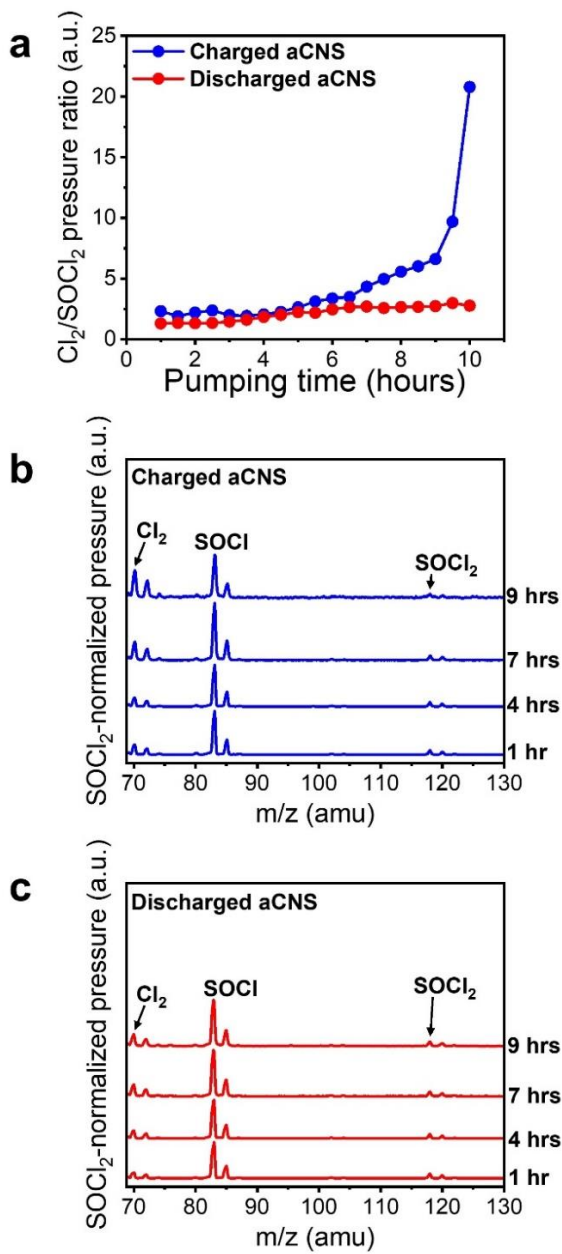

**Figure 4. Mass-spectrometry studies of charged and discharged aCNS electrodes. a,** Detected pressure ratio between $Cl_2$ (m/z = 70 amu) and $SOCl_2$ (m/z = 118 amu) at different pumping times. This ratio increased as pumping time increased in charged aCNS electrode, suggesting that free $Cl_2$ was gradually released from the electrode. This ratio remained roughly constant as pumping time increased in discharged aCNS electrode, suggesting that all the $Cl_2$ signal detected was due to fragmentations of $SOCl_2$. **b,** Recorded

SOCl$_2$-normalized mass spectra (SOCl$_2$ intensity = 1) of charged aCNS electrode at different pumping times. The pumping time was labeled next to each spectrum. The recorded Cl$_2$ intensity clearly increased as pumping time increased, suggesting the releasing of free Cl$_2$ from the electrode. **c,** Recorded SOCl$_2$-normalized mass spectra (SOCl$_2$ intensity = 1) of discharged aCNS electrode at different pumping times. The pumping time was labeled next to each spectrum. The recorded Cl$_2$ intensity remained roughly the same as pumping time increased, suggesting that Cl$_2$ came from the fragmentations of SOCl$_2$.

Lastly, we performed ex situ XPS measurements of charged (to 1200 mAh g$^{-1}$) and discharged aCNS without any exposure to air or water (see Methods), within two hours of sample transferring to the XPS vacuum system. From mass spectrometry, detectable Cl$_2$ remained in the carbon electrode within several days of vacuuming at room temperature (Fig. 4). The Cl 2p XPS features of the discharged aCNS showed NaCl (Cl 2p$_{3/2}$ = ~ 198.6 eV, formed during the first battery discharge, not completely re-oxided, and formed during cycling[1]) and AlCl$_3$ (Cl 2p$_{3/2}$ = ~ 199.4 eV, part of the electrolyte, Fig. 5a) on the electrode. Upon charging to 1200 mAh g$^{-1}$, the relative peak intensity of the NaCl component decreased, consistent with electro-oxidation of NaCl. Interestingly, a new peak at ~ 200 eV appeared, discernable in both the raw XPS spectrum and after peak fitting (Fig. 5b). Based on the relative sensitivity factors (RSF) of Cl 2p and C 1s, we estimated that the detected Cl 2p peak at ~ 200 eV constituted ~ 7.7% of the total intensity of C 1s, which was about twice of the Cl 2p peak intensity at ~ 200 eV assigned to C-Cl for water washed electrode (Fig. 2). Hence, we attributed the strong ~ 200 eV peak component (Fig. 5b) largely to Cl$_2$ remaining trapped in the vacuumed aCNS (in addition to Cl in C-Cl bonds). Previously there was only one report on Cl 2p$_{3/2}$ peak at ~ 200 eV for molecular Cl$_2$ (~ 2.5 monolayers) adsorbed on a Ag/AgCl surface probed by synchrotron-based XPS at 100 K[36]. We also performed XPS measurement of a 1200 mAh g$^{-1}$ charged aCNS electrode after vacuuming the electrode for 96 hours and observed much reduced intensity of the ~ 200 eV peak (Fig. 5c, indicated by '*'), suggesting removal of $Cl_2$ trapped in aCNS with the remaining signals corresponding to Cl in C-Cl bonding that had similar concentration (~ 3.81%) as in DIUF-water-washed aCNS (~ 3.87%, Fig. 2b). To the best of our knowledge, this was the first time that $Cl_2$ was detected by XPS in vacuum at room temperature, owing to strong trapping of $Cl_2$ in the porous carbon nanosphere.

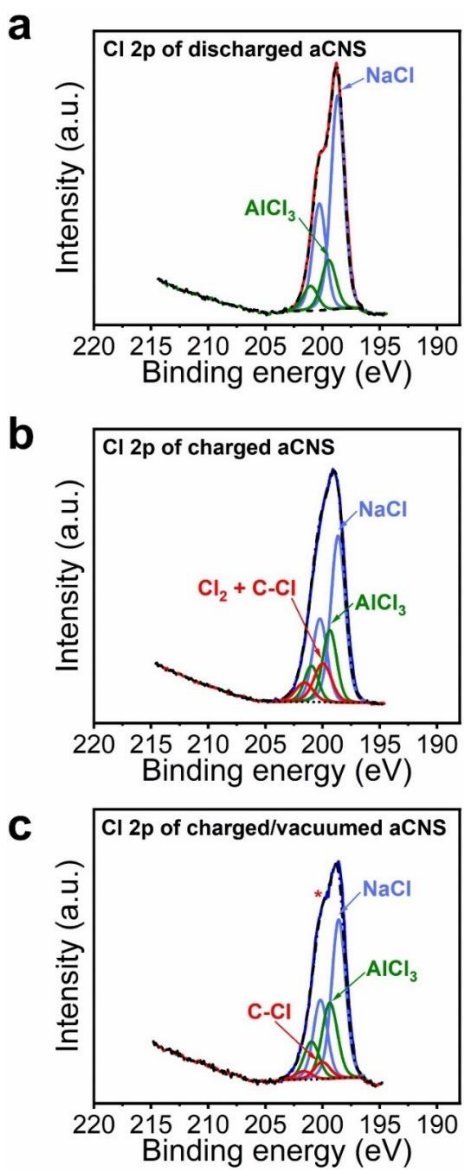

**Figure 5. XPS studies of aCNS electrodes at different states during battery operation without exposing these electrodes to air or water. a,** Cl 2p XPS spectrum of aCNS electrode after the battery was fully discharged. The spectrum was well-fitted using

only 2 sets of doublet, one for NaCl and one for $AlCl_3$. **b,** Cl 2p XPS spectrum of aCNS electrode after the battery was charged to 1200 mAh $g^{-1}$. In addition to the peaks due to NaCl and $AlCl_3$, a new peak at ~ 200 eV, attributed to a mixture of $Cl_2$ and C-Cl, was needed to fit the overall spectrum, suggesting that $Cl_2$ was formed during battery charging. **c,** Cl 2p XPS spectrum of aCNS electrode after the battery was charged to 1200 mAh $g^{-1}$ (same sample as b) but pumped under vacuum for 96 hours. After pumping, in addition to the peaks due to NaCl and $AlCl_3$, an obvious decrease in the intensity of the ~ 200 eV peak was observed (indicated by '*'), suggesting that most of the $Cl_2$ was vacuumed away during the pumping period. The XPS fitting was done using the CasaXPS software with a Lorentzian asymmetric line shape. All the peaks used in fitting had their binding energy and FWHM restrained to vary by ± 0.1 eV.

**Conclusion**

We performed in-depths spectroscopic investigations of rechargeable $Na/Cl_2$ battery. We found that during battery charging, aCNS was chlorinated to form C-Cl bonds, accompanied by infiltration of $Cl_2$ into the micropores of the carbon matrix, and the development of graphitic ordering in the initially amorphous aCNS with intercalations of chloride-based anionic species into the nanoscale graphitic regions. During battery discharging, these effects were reversed with de-chlorination of carbon, de-intercalation and reversal of graphitic ordering to a full amorphous state. In addition, by employing mass spectrometry, we probed molecular $Cl_2$ trapped in the porous aCNS in charged state and observed $Cl_2$ trapped in the charged electrode under vacuum by XPS. Overall, this work shed more light on the carbon positive electrode evolution during $Na/Cl_2$ battery cycling, and confirmed conversion between $Cl_2$ and NaCl in the carbon matrix as a main redox reaction of the rechargeable $Na/Cl_2$ battery.

**Methods**

**Synthesis of aCNS and fabrication of aCNS electrode**

aCNS was synthesized following the same method as reported in our previous paper[1]. After aCNS was synthesized, 90% by weight of aCNS powder was mixed with 10% by weight of polytetrafluoroethylene (PTFE, 60% aqueous dispersion, FuelCellStore). 100% ethanol (Fisher Scientific) was then added to the mixture and the mixture was sonicated for 2 hours. Nickel (Ni) foam or SS316 stainless steel mesh was used as the electrode substrate and was cut into circular pieces with 1.5 cm in diameter using a compact precision disk cutter (MTI, MSK-T-07). The substrates were then sonicated in 100% ethanol for 15 minutes and dried at 80 °C until all the ethanol was evaporated. The circular substrate was then weighed and hovered over a hot plate. The mixture of aCNS, PTFE, and ethanol was then slowly dropped onto the substrate (180 μL each time). The ethanol from previous drop must be completely evaporated before the next drop was added to the substrate. The process was stopped until a desirable amount of aCNS was loaded on the substrate, after which the aCNS electrode was dried at 80 °C overnight. The electrode was then pressed using a spaghetti roller and its final weight was also measured and recorded. The final weight of aCNS was determined by the final weight of the electrode minus the initial weight of the substrate times 90%. After the weight of aCNS was determined, the electrode was stored in 80 °C oven until being used in a Na/$Cl_2$ battery.

**Electrolyte making**

Electrolyte was made inside an argon-filled glovebox. Thionyl chloride ($SOCl_2$, Spectrum Chemical Mfg. Corp. TH138-100ML) was purchased and used without any further purification. The appropriate amount of $SOCl_2$ was transferred into a 20-mL scintillation vial (Fisher Scientific) with the weight of $SOCl_2$ being measured and recorded. 4 M $AlCl_3$ (Fluka, 99%, anhydrous, granular) was weighed and added to the $SOCl_2$. The mixture was then stirred until all the $AlCl_3$ was dissolved. NaFSI (TCI America) and NaTFSI (TCI America) were stored inside a 90 °C vacuum oven before being used. After all the $AlCl_3$ was dissolved, 2% by weight, i.e., weight of $SOCl_2$ plus weight of $AlCl_3$ times 2%, of NaFSI (TCI America) and NaTFSI (TCI America) were added to the mixture and stirred until all the chemicals were dissolved. The electrolyte was then ready to be used in a $Na/Cl_2$ battery.

**Battery making**

All batteries were assembled inside an argon-filled glovebox. Sodium metal block (Sigma Aldrich) was taken out from the mineral oil and the oil was cleaned using kimwipes (Kimberly-Clark Professional™ Kimtech Science™). A razor blade was used to cut small pieces from all sides of the sodium metal block so that the shiny sodium metal was exposed. The shiny sodium metal block was then placed inside a zip lock bag (Ziploc) and pressed with a scintillation vial so that the metal block became a thin sodium foil. The thin sodium foil was then pasted onto the coin cell spacer (MTI corporation) and any extra sodium foil was removed. In the end, the sodium metal should have the same size and shape as the coin cell spacer and completely cover the spacer. After the sodium metal negative electrode was made, aCNS electrode was then placed in the center of the positive coin cell case (MTI corporation, SS316, CR2032). Two layers of quartz fiber filters (Sterlitech, QR-100) was then placed on top of the aCNS electrode as the separators. 150 μL of electrolyte was then added to the separators and the sodium metal negative electrode was placed directly facing the aCNS electrode. Coin cell spring (MTI corporation) was placed on top of the spacer and finally a negative coin cell case was used to cover the entire battery. The battery was then sealed using a digital pressure controlled electric crimper (MTI corporation, MSK-160E) with the pressure reading at 13.2. After the battery was sealed, it was taken out from the glovebox and a layer of GE advanced silicone sealant was applied to cover the O-ring of the battery at which the two cases were sealed. The purpose of this layer of silicone was to further protect the battery from leaking so that it would not react with air and water. After the silicone was cured (~ 25 minutes), the battery was tested using a Neware battery tester (Neware, CT-4008-5V50mA-164-U).

**Synchrotron XRD measurement**

The two-dimensional x-ray diffraction (2D-XRD) measurements for aCNS loaded on Ni foam substrate at different states during battery cycling were performed at the BL01C2 in the Taiwan Light Source (TLS) of National Synchrotron Radiation Research Center (NSRRC), Hsinchu, Taiwan. The beamline was operated at the energy of 16 KeV. The XRD pattern was recorded using the λ=0.7749 Å with 0.9 cm × 0.2 cm focus beam spot size in the 2θ ranging from 5º-39º by image plate detector mar345s. Each sample was exposed for 30 seconds and was calibrated by standard sample of $CeO_2$. The aCNS sample was taken out from an opened battery and was sealed in the CR2032 coin cell case with a 5 mm hole covered by Kapton and epoxy glue inside an argon-filled glovebox with both $H_2O$ and $O_2$ levels of < 1 ppm. All the crystallography data of the 1D, 2D XRD pattern was analyzed by GSAS-II. The XRD pattern reported in this work had its x-ray wavelength converted to that of copper K-α (1.5406 Å).

**XPS measurement**

XPS measurements were conducted in Stanford Nano Shared Facilities (SNSF) using a PHI VersaProbe 3 instrument.

To acquire the XPS spectra of DIUF-water-washed aCNS electrodes, the $Na/Cl_2$ battery with SS316 stainless steel mesh as the positive electrode substrate was opened inside an argon-filled glovebox immediately after the battery had reached its designated cycling state (1200 mAh $g^{-1}$ charged or fully discharged). The aCNS electrode (together with a small amount of separator stuck to its surface) was then taken out from the opened battery and put into a 20-mL scintillation vial. The vial was then transferred outside the glovebox and DIUF water was added into the vial. The electrode inside the vial was stirred for ~ 2 minutes, and the DIUF water was then poured out from the vial. New DIUF water was then added to the vial and the process was repeated five times. During this process, the separator that was originally stuck to the electrode would peel off, allowing some aCNS powder to fall off together with the separator. After the washing process was completed, the separator with aCNS powder stuck to it was then transferred inside a vacuum chamber for drying. After the sample was dried, it was transferred inside the PHI VersaProbe 3 instrument in SNSF and its XPS spectrum was recorded.

To acquire the XPS spectra of aCNS electrodes without exposing them to air or water, a special vacuum transfer vessel (VTV, Physical Electronics) was used. The VTV and the platen for XPS instrument were first transferred together into an argon-filled glovebox. After the $Na/Cl_2$ battery with SS316 stainless steel mesh as the positive electrode substrate had cycled to its designated state, the battery was disassembled inside the argon-filled glovebox, and the aCNS electrode was then taken out from the battery and placed onto the platen used for the XPS instrument. The platen was then put inside the VTV, which was then completely sealed in the glovebox. After the VTV was sealed, it was transferred out from the glovebox and installed to the intro chamber of the PHI VersaProbe 3 instrument in SNSF. The VTV was then slowly opened, and the sample was able to be transferred into the instrument for XPS measurements.

To measure the XPS of aCNS after annealing at 600 °C in a $N_2$ environment. The DIUF-water-washed aCNS powder stuck to the separator was first obtained (see above). Then the sample was transferred inside a quartz tube, which was then put onto a tube furnace. Nitrogen gas was allowed to flow through the quartz tube at a flow rate of 100 sccm. The temperature was then gradually increased to 600 °C at a rate of 10 °C min$^{-1}$ and remained at this target temperature for 45 minutes. Afterwards, the furnace was allowed to cool down naturally, and the aCNS sample was taken out from the quartz tube. Then the sample was transferred inside the PHI VersaProbe 3 instrument in SNSF for XPS measurements.

**AES with scanning Auger nanoprobe measurement**

The AES and elemental mapping were conducted in SNSF using a PHI 700 scanning Auger nanoprobe. The Na/Cl$_2$ battery with Ni foam as the positive electrode substrate was disassembled inside an argon-filled glovebox once it had reached its designated cycling state. The aCNS electrode was then taken out from the opened battery and well-sealed inside an aluminum laminated pouch (MTI, EQ-alf-100-210). The pouch was then taken to SNSF and opened immediately before the sample was ready to be transferred into the instrument for measurements. During the transfer process, the sample's exposure time to air was minimized (< 20 seconds).

**Mass spectrometry measurement**

Mass spectrometry studies were done using a residual gas analyzer (Fig. S4, RGA-300, Stanford Research Systems). The Na/Cl$_2$ battery with Ni foam as the positive electrode substrate was disassembled inside an argon-filled glovebox once it had reached its designated cycling state. The aCNS electrode with a small amount of separator stuck to its surface was then taken out from the battery and immediately transferred into the sample chamber with valve #2 closed (Fig. S4). The sample chamber was then taken out from the glovebox and connected to the instrument via the capillary tube at the connection point (Fig. S4). With valve #2 closed, valve #1 was then opened and the entire system was pumped for 1 hour to remove any residual air inside. After 1 hour of pumping, valve #2 was opened, and species inside the sample chamber were continuously pumped to the instrument for mass spectrometry measurements. The mass spectra at different pumping times were then recorded and reported in this work.

**Acknowledgements**

This work was funded by the Deng family gift. Part of this work was performed at the Stanford Nano Shared Facilities (SNSF), supported by the National Science Foundation under award ECCS-2026822.

**Author contributions**

G. Zhu and H. Dai conceived the main idea of this work. G. Zhu and P. Liang performed the experiments and contributed equally to this work. G. Zhu assembled all the batteries used in this work. G. Zhu and P. Liang performed the XPS and mass spectrometry analysis. G. Zhu and S. -C. Wu performed charge-discharge of the batteries. G. Zhu performed the AES mapping measurement. C. -L. Huang, C. -C. Huang, and Y. -Y. Li prepared aCNS and performed characterizations on the material. S. -K. Jiang, W. -H. Huang and B. -J. Hwang performed the XRD characterizations on aCNS. J. Li and F. Wang performed the Raman measurement on aCNS. G. Zhu and H. Dai prepared the manuscript. All authors participated in experimental data/results analysis and discussion.

# Supporting Information

## Shedding Light on Rechargeable Na/Cl$_2$ Battery

Guanzhou Zhu[1§], Peng Liang[1§], Cheng-Liang Huang[2,3], Shu-Chi Wu[1], Cheng-Chia Huang[2], Yuan-Yao Li[2], Shi-Kai Jiang[4], Wei-Hsiang Huang[5], Jiachen Li[1], Feifei Wang[6], Bing-Joe Hwang[4], Hongjie Dai[1*]

[1]Department of Chemistry and Bio-X, Stanford University, Stanford, California 94305, USA.

[2]Department of Chemical Engineering, National Chung Cheng University, Chia-Yi 62102, Taiwan.

[3]Department of Electrical Engineering, National Chung Cheng University, Chia-Yi 62102, Taiwan.

[4]Department of Chemical Engineering, National Taiwan University of Science and Technology, Taipei 10607, Taiwan.

[5]National Synchrotron Radiation Research Center (NSRRC), Hsinchu 30076, Taiwan.

[6]Department of Electrical and Electronic Engineering, The University of Hong Kong, Hong Kong 999077, Hong Kong.

§ G.Z. and P.L. contributed equally to this work.

* Email: hdai1@stanford.edu.

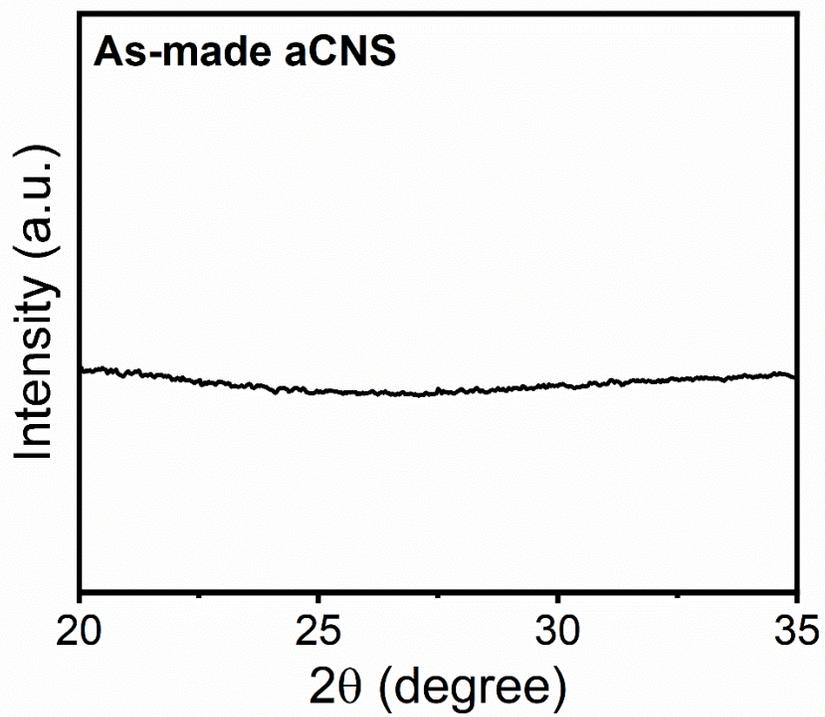

**Figure S1. XRD spectrum of as-made aCNS in 2θ ranging from 20º to 35º.** No diffraction peak was observed, indicating that aCNS was indeed amorphous in nature.

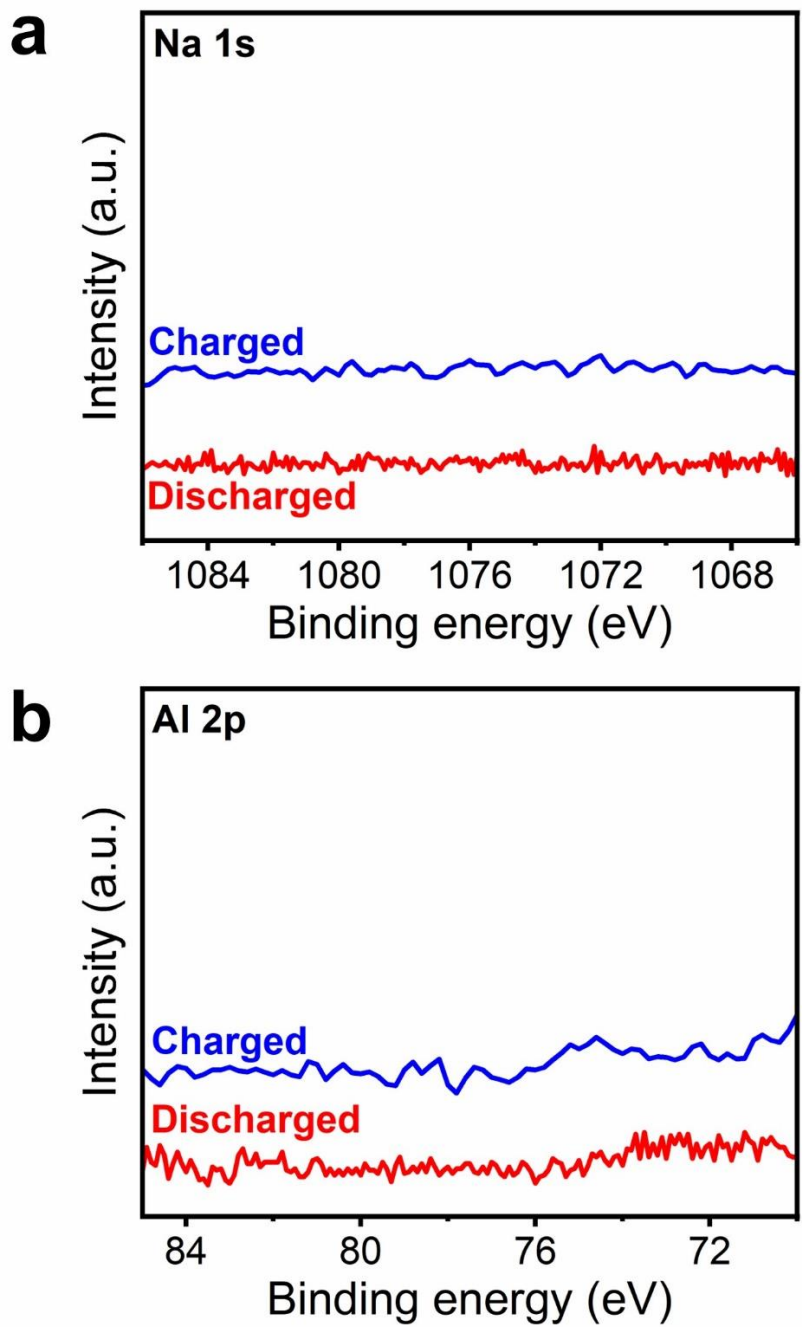

**Figure S2. Na 1s and Al 2p XPS spectra of charged and discharged aCNS electrodes after washing using DIUF water. a,** Na 1s spectra of charged aCNS (blue curve) and discharged aCNS (red curve). No obvious Na 1s peak was observed, indicating that all the Na salts in the electrode were washed away by DIUF water. **b,** Al 2p spectra of charged aCNS (blue curve) and discharged

aCNS (red curve). No obvious Al 2p peak was observed, indicating that all the Al salts in the electrode were washed away by DIUF water.

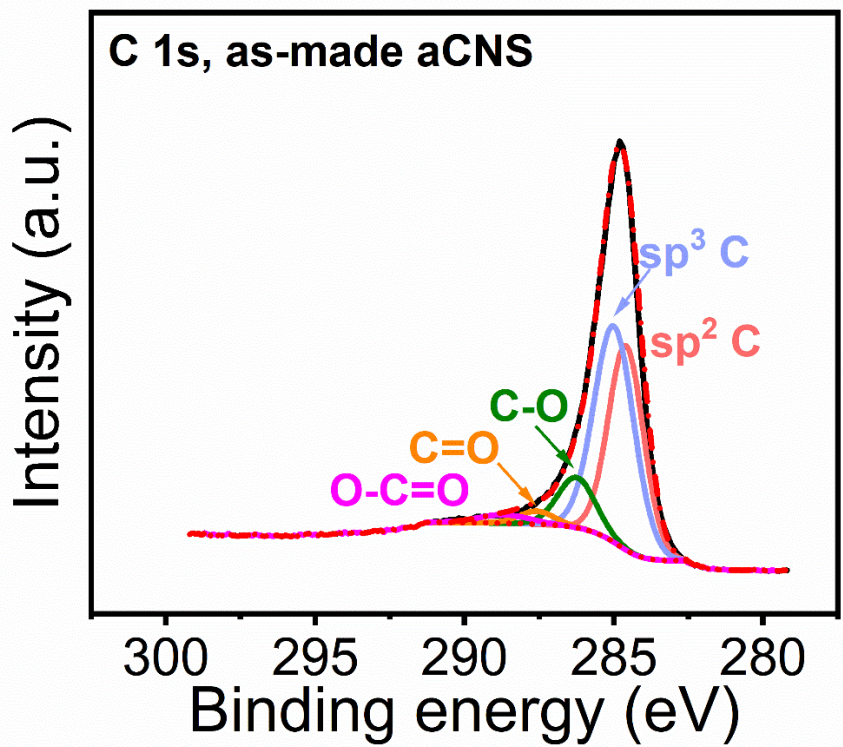

**Figure S3. C 1s XPS spectrum of the as-made aCNS.** The C 1s spectrum can be well fitted using the following few peaks: $sp^2$ C (~ 284.5 eV), $sp^3$ C (~ 285.0 eV), C-O (~ 286.3 eV), C=O (287.7 eV), and O-C=O (~ 288.7 eV).

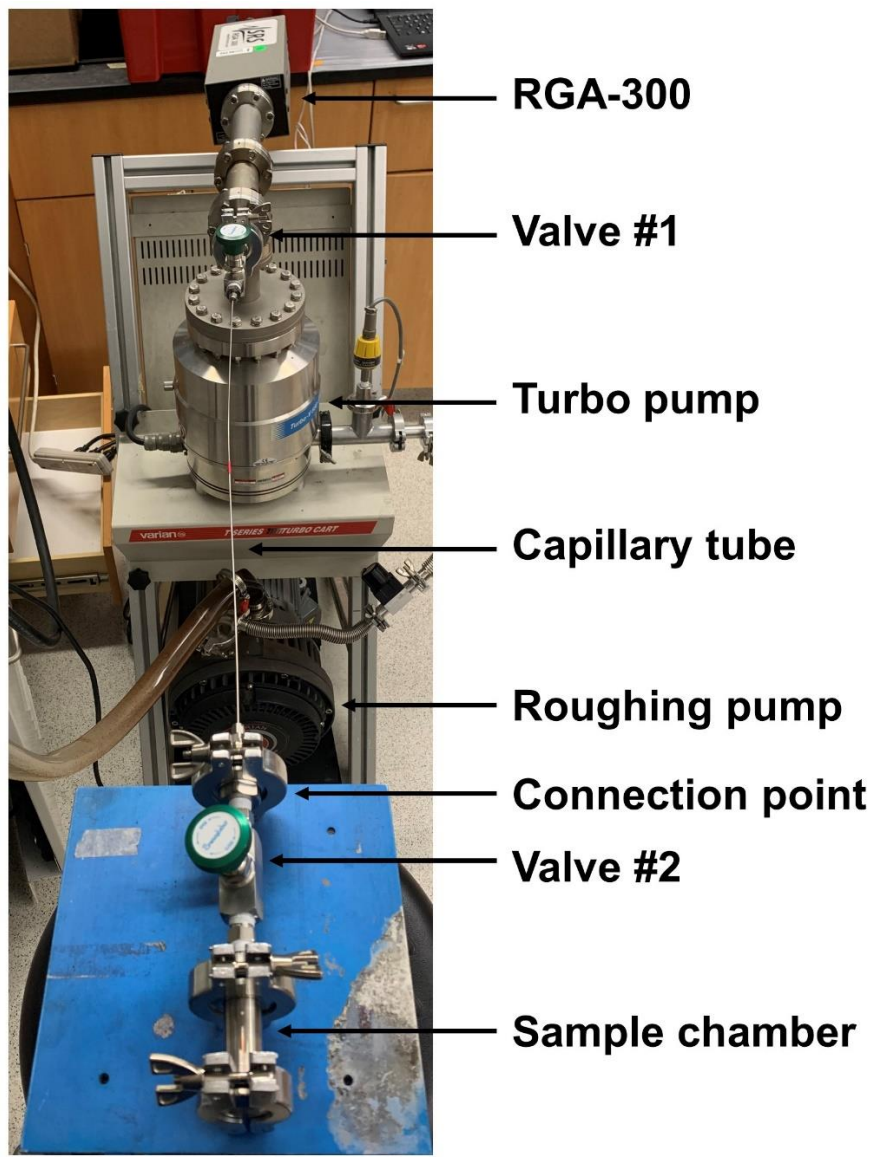

**Figure S4. Experimental setup of our residual gas analyzer that was used for mass spectrometry studies.** See Methods for details.